\documentclass[aps,prd,floats,floatfix,twocolumn,showpacs,superscriptaddress,nofootinbib]{revtex4}
\usepackage{amsmath, amsfonts}
\usepackage{graphicx}

\newcommand{\SMetric}{g}     
\newcommand{\Lapse}{N}       
\newcommand{\Shift}{\beta}   

\newcommand{\ExCurv}{K}      
\newcommand{\TrExCurv}{K}    

\newcommand{\dtime}{\partial_t}   

\newcommand{\CF}{\psi}              

\newcommand{\CMetric}{{\tilde{g}}}    
\newcommand{\CLapse}{\tilde{N}}     
\newcommand{\dtCMetric}{\tilde{u}}  

\newcommand{\CA}{\tilde{A}}         
\newcommand{\CRicciS}{\tilde{R}}    

\newcommand{\CCD}{{\tilde\nabla}\!} 
\newcommand{\CCDu}{{\tilde\nabla}}  
\newcommand{\CLong}[1]{(\tilde{\mathbb L}{#1})} 

\begin{document}

\title{The Einstein constraints: uniqueness and non-uniqueness in the
conformal thin sandwich approach}

\author{Thomas W. Baumgarte}
\email{tbaumgar@bowdoin.edu}
\affiliation{Department of Physics and Astronomy, Bowdoin College,
Brunswick, ME 04011, USA}
\altaffiliation{Also at Department of Physics, University of Illinois,
Urbana, Il 61801}
\author{Niall \'{O} Murchadha}
\email{niall@ucc.ie}
\affiliation{Physics Department, University College, Cork, Ireland}

\author{Harald P. Pfeiffer}
\email{harald@tapir.caltech.edu}
\affiliation{Theoretical Astrophysics, California Institute of Technology, 
Pasadena, CA 91125, USA}


\begin{abstract}
We study the appearance of multiple solutions to certain
decompositions of Einstein's constraint equations.  Pfeiffer and York
recently reported the existence of two branches of solutions for
identical background data in the extended conformal thin-sandwich
decomposition.  We show that the Hamiltonian constraint alone, when
expressed in a certain way, admits two branches of solutions with
properties very similar to those found by Pfeiffer and York.  We
construct these two branches analytically for a constant-density star
in spherical symmetry, but argue that this behavior is more general.
In the case of the Hamiltonian constraint this non-uniqueness is well
known to be related to the sign of one particular term, and we argue
that the extended conformal thin-sandwich equations contain a similar
term that causes the breakdown of uniqueness.
\end{abstract}

\maketitle

\section{Introduction}

With the help of a 3+1 decomposition Einstein's equations can be split
into a set of constraint equations and a set of evolution equations
\cite{Arnowitt-Deser-Misner:1962,York:1979}.  The four constraint
equations -- one in the Hamiltonian constraint and three in the
momentum constraint -- constrain the induced spatial metric
$\SMetric_{ij}$ and the extrinsic curvature $\ExCurv_{ij}$ on spatial
hypersurfaces representing instants of constant coordinate time $t$.
The constraint equations constrain only four of these initial
variables; the remaining ones are freely specifiable and have to be
chosen independently before the constraint equations can be solved.  A
decomposition of the initial data separates the freely specifiable
variables from the constrained ones.  Given a particular
decomposition, the construction of initial data then entails making
well-motivated choices for the freely specifiable independent
background data and then solving the constraint equations for the
constrained variables.
 
The conformal thin-sandwich decomposition has emerged as a
particularly popular decomposition among numerical relativists,
especially for the construction of quasi-equilibrium data (see, e.g.,
the reviews \cite{Cook:2000,Baumgarte-Shapiro:2003} and a brief
discussion below).  This variation of the original (non-conformal)
thin-sandwich decomposition
\cite{Belasco-Chanian:1969,Bartnik-Fodor:1993,Giulini:1999} was
formally developed by York \cite{York:1999} (see also
\cite{Pfeiffer-York:2003}), but more restricted versions had been
introduced earlier \cite{Isenberg:1978,Wilson-Mathews:1989}.  It has
been used, for example, for the construction of binary neutron stars
(e.g.
\cite{Baumgarte-Cook-etal:1997,Bonazzola-Gourgoulhon-Marck:1999,
Uryu-Eriguchi:2000,Gourgoulhon-Grandclement-etal:2001}), binary black
holes (e.g.  \cite{Gourgoulhon-Grandclement-Bonazzola:2001a,
Grandclement-Gourgoulhon-Bonazzola:2001b,Yo-Cook-etal:2004,
Cook-Pfeiffer:2004,Caudill-Cook-etal:2006}) and black hole-neutron star 
binaries (\cite{Baumgarte-Skoge-Shapiro:2004,Taniguchi-Baumgarte-etal:2005}).

Given this wealth of experience with the conformal thin-sandwich
decomposition, it came as quite a surprise when Pfeiffer and York
(\cite{Pfeiffer-York:2005}, hereafter PY) recently discovered
non-uniqueness in the solution of the conformal thin-sandwich system.
Even for `small' independent background data, which one would expect
to generate gravitational initial data close to a flat slice of flat
spacetime, the so-called ``extended'' set of conformal thin-sandwich
data allowed for two branches of solutions.  One of these two branches
has comparatively weak gravitational fields and indeed approaches flat
space in the limit of vanishing background data, while the second
strong-field branch approaches a singular solution.

In this paper we discuss some of the aspects of this non-uniqueness.
We consider a spherically symmetric, constant density star and show
that even for this very simple model we can find -- for sufficiently
small values of the density -- two branches of solutions.  These two
branches share some of the characteristics of the solutions found by
PY and illustrate their properties.  For our spherically symmetric
solution the non-uniqueness of solutions is caused by a particular
term having the ``wrong sign'' (see, e.g.~\cite{York:1979}), and we
argue that the non-uniqueness found in the extended conformal
thin-sandwich equations may be caused by a similar term.  

We note also that certain constrained evolution
schemes~\cite{Choptuik-Hirschmann-Liebling-Pretorius:2003,Rinne:2005}
solve a set of elliptic equations at every timestep which is very
similar to the extended conformal thin sandwich equations.  These
authors observed occasional failure of their elliptic solvers in the
strong field regime, and it was
argued~\cite{Rinne:2005,Rinne-Steward:2005} that this failure is
caused by the ``wrong sign'' in the maximal slicing condition.

This paper is organized as follows.  In Section \ref{Sec:CTS} we
briefly review the conformal thin-sandwich decomposition.  We then
consider the Hamiltonian constraint in spherical symmetry in Section
\ref{Sec:HamToy}.  First, in Sec.~\ref{SubSec:ToyModel}, we
construct analytic solutions for constant density stars and show that
these solutions consist of two branches with properties very similar
to the solutions found by PY.  Subsequently, in
Sec.~\ref{SubSec:Math}, we prove that at least some of these
properties persist for arbitrary spherically symmetric solutions.
Finally, we provide a brief summary and discussion in Section
\ref{Sec:Sum}.
 
\section{The conformal thin-sandwich decompositions}
\label{Sec:CTS}

Conformal decompositions of the constraint equations start with a
conformal transformation of the spatial metric, $\SMetric_{ij} = \CF^4
\CMetric_{ij}$, where $\CF$ is the conformal factor and
$\CMetric_{ij}$ the conformally related metric.  The Hamiltonian
constraint then becomes an equation for the conformal factor
\begin{equation}
\label{eq:Ham2}
\CCDu^2\CF-\frac{1}{8}\CRicciS\CF-\frac{1}{12}\TrExCurv^2\CF^5 
+\frac{1}{8}\CF^{-7}\CA^{ij}\CA_{ij} + 2\pi \CF^5 \rho = 0.
\end{equation}
Here $\CCD$ and $\CRicciS$ are the covariant derivative and the trace of
the Ricci tensor associated with $\CMetric_{ij}$, and the extrinsic
curvature is decomposed into its trace $K$ and the conformally related
trace-free part $\CA^{ij}$,
\begin{equation}
K^{ij} =  \psi^{-10} \CA^{ij} + \frac{1}{3} g^{ij} K.
\end{equation}
For completeness we have also included the matter source $\rho = n^a
n^b T_{ab}$, where $n^a$ is the normal on the spatial hypersurface and
$T_{ab}$ the stress-energy tensor, and where summation is carried out
over four spacetime indices. 

The matter term as written in Eq.~({\ref{eq:Ham2}) has the
defect~\cite{York:1979} that its positive sign combined with the
positive exponent of $\CF$ {\em prevent} use of the maximum principle
to prove local uniqueness of solutions.  Therefore, it is not
immediately clear that solutions to Eq.~(\ref{eq:Ham2}) are unique
(indeed, we show in Sec.~\ref{Sec:HamToy}, that often they are not
unique).  This defect can be cured~\cite{York:1979} by introduction of
a conformally scaled matter density $\tilde\rho=\CF^{8}\rho$; taking
$\tilde\rho\ge 0$ as freely specifiable data, the matter term in
Eq.~({\ref{eq:Ham2}) becomes $2\pi\CF^{-3}\tilde\rho$. Because of the
sign-change in the exponent, this term is now well-behaved and the
maximum principle is applicable.  We will use Eq.~(\ref{eq:Ham2}) with
the unscaled $\rho$ as a toy example in Sec.~\ref{Sec:HamToy} below.
Besides that, we are only interested in vacuum space-times and
therefore do not include matter terms in the rest of this Section.

The conformal metric $\CMetric_{ij}$, meanwhile, is freely
specifiable. In the conformal thin-sandwich decompositions, the time
derivative of the conformal metric, $\dtCMetric_{ij} \equiv
\dtime\CMetric_{ij}$ is also considered freely specifiable.  Using the
evolution equation for the spatial metric we can relate
$\dtCMetric_{ij}$ to $\CA_{ij}$,
\begin{equation}\label{A}
\CA^{ij} =
\frac{1}{2\CLapse}\Big(\CLong{\Shift}^{ij}-\dtCMetric^{ij}\Big),
\end{equation}
where the conformal (or densitized) lapse $\CLapse$ is related to the
lapse $N$ by $\Lapse = \CF^6 \CLapse$.  Inserting this expression into
the momentum constraint yields 
\begin{equation}\label{eq:Mom2}
\CCD_j\Big(\frac{1}{2\CLapse}\CLong{\Shift}^{ij}\Big)
-\frac{2}{3}\CF^6\CCDu^i\TrExCurv
-\CCD_j\Big(\frac{1}{2\CLapse}\dtCMetric^{ij}\Big) = 0
\end{equation}
where $\CLong{\Shift}^{ij}\equiv 2\CCDu^{(i}\Shift^{j)}
-2/3\,\CMetric^{ij}\CCD_k\Shift^k$ is the conformal longitudinal
operator.

There are two versions of the conformal thin sandwich approach. In the
{\em standard} conformal thin sandwich equations, one specifies
$(\CMetric_{ij}, \dtCMetric_{ij}; \TrExCurv, \CLapse)$ and suitable
matter-terms, if applicable.  Given these background variables,
Eq.~(\ref{eq:Ham2}) and (\ref{eq:Mom2}) (together with (\ref{A})) can
be solved for the conformal factor $\CF$ and the shift $\Shift^i$,
which completes the set of initial data.  

For maximal slices, $\TrExCurv=0$, Eqs.~(\ref{eq:Ham2}) and
(\ref{eq:Mom2}) decouple, so that Eq.~(\ref{eq:Mom2}) can be
considered first.  For any given strictly positive $\CLapse$, this
equation is a linear elliptic equation so that the existence of a
unique solution $\Shift^i$ is guaranteed. This is a key motivation for
the entire structure and is discussed in
\cite{York:1999,Pfeiffer-York:2003}.  Eq.~(\ref{eq:Ham2}) -- with
zero matter density $\rho$ -- becomes the
standard Lichnerowicz equation for the conformal factor
\cite{Lichnerowicz:1944} and again has a unique solution as long as
the base metric is in the positive Yamabe class.

In the {\em extended} system one regards $\dtime\TrExCurv$
instead of $\CLapse$ as freely specifiable.  The lapse can then be
solved for from the trace of the evolution equation for the extrinsic
curvature, which often is written as 
\begin{align}
\nonumber
\CCDu^2(\CLapse\CF^7)-(\CLapse\CF^7)&\bigg[\frac{\CRicciS}{8}\!+\!\frac{5}
{12}\TrExCurv^4\CF^4\!
+\!\frac{7}{8}\CF^{-8}\CA^{ij}\CA_{ij}\bigg]\\
\label{eq:Lapse2}
&=-\CF^5(\dtime\TrExCurv-\Shift^k\partial_k\TrExCurv).
\end{align}
The independent background data now
are $(\CMetric_{ij}, \dtCMetric_{ij}; \TrExCurv, \partial_t\TrExCurv)$
(and suitable matter terms, if applicable) and we solve five coupled
elliptic equations (\ref{eq:Ham2}), (\ref{eq:Mom2}) and
(\ref{eq:Lapse2}) for the conformal factor $\CF$, the shift $\Shift^i$
and the lapse $N$.  This extended system has become very popular in
numerical relativity because the ability to set the time derivatives
$\dtCMetric_{ij}$ and $\dtime\TrExCurv$ to zero provides a means of
constructing quasi-equilibrium data.

However, PY demonstrated that the extended conformal thin sandwich
equations behave very differently from the standard set, even for
$\TrExCurv=0=\dtime\TrExCurv$.  Specifically, they found two branches
of solutions for the same choices of free data.  

We wish to point out that Eq.~(\ref{eq:Lapse2}) is
written in a misleading way. As written, it appears that
the maximum principle can be used for Eq.~(\ref{eq:Lapse2}).
However, $\tilde A^{ij}$ contains the lapse itself, cf. Eq.~(\ref{A});
displaying this dependence explicitly results in
\begin{align}
\nonumber
\CCDu^2(&\CLapse\CF^7)-\frac{7}{32}\frac{\CF^{6}}{(\CLapse\CF^7)}
\left(\CLong\Shift^{ij}-\dtCMetric^{ij}\right)
\left(\CLong\Shift_{ij}-\dtCMetric_{ij}\right)
\\
&-(\CLapse\CF^7)\bigg[\frac{1}{8}\CRicciS\!+\!\frac{5}
{12}\TrExCurv^4\CF^4\!\bigg]=
-\CF^5(\dtime\TrExCurv-\Shift^k\partial_k\TrExCurv).
\label{eq:Lapse3}
\end{align}
The first line of this equation has the structure
\begin{equation*}
\CCDu^2(\CLapse\CF^7)-f \left(\CLapse\CF^7\right)^{-1}
\end{equation*}
with non-negative coefficient $f$.  The sign of $f$ combined with the
negative exponent of $(\CLapse\CF^7)$ in the second term prevents
application of the maximum principle, as did the unscaled density 
term in the Hamiltonian constraint (\ref{eq:Ham2}).  

We believe that this term might very well be responsible for the
complex behavior exhibited by the extended conformal thin sandwich
equations.  To support our claim, we analyze the Hamiltonian
constraint (\ref{eq:Ham2}) with an unscaled density in Section
\ref{Sec:HamToy} below.  We construct an analytic solution in
spherical symmetry and explicitly show the existence of two branches
of solutions with properties very similar to those reported by PY.



\section{Hamiltonian constraint with unscaled matter density}
\label{Sec:HamToy}

As we discussed above, the Hamiltonian constraint Eq.~(\ref{eq:Ham2})
with {\em unscaled} matter density is not amenable to the maximum
principle, and it turns out to be interesting investigate consequences
of this fact.  We consider the initial value problem at a moment of
time-symmetry, $\ExCurv_{ij}\equiv 0$, so that the momentum constraint
is satisfied identically.  Assuming further conformal flatness and
spherical symmetry, the Hamiltonian constraint Eq.~(\ref{eq:Ham2})
reduces to
\begin{subequations}
\label{eq:Toy1}
\begin{equation}\label{eq:Toy1a}
\nabla^2\CF+2\pi\rho\CF^5=0,\\
\end{equation}
with $\CF>0$ and with boundary conditions
\begin{align}\label{eq:Toy1b}
\frac{\partial\CF}{\partial r}=0,&\qquad r=0,\\
\CF\to 1,&\qquad r\to\infty,
\end{align}
\end{subequations}
where $\nabla^2=\partial^2/\partial r^2+(2/r)\partial/\partial r$
represents the flatspace Laplacian, and we assume a density profile
$\rho(r)\ge 0$.

\subsection{The constant density star}
\label{SubSec:ToyModel}

First we will consider a constant density star of (conformal) radius
$R$ and mass-density
\begin{equation}\label{eq:Toy2}
\rho(r)=\left\{\begin{aligned}&\rho_0,&r<R,\\
&0,&r>R.
\end{aligned}
\right.
\end{equation}
We will take $R$ to be fixed, and examine the solutions of this
equation as we vary $\rho_0$.  Thus, $\rho_0$ plays the role of the
``amplitude'' of the perturbation away from trivial initial data.

Solutions of Eq.~(\ref{eq:Toy1}) in the interior of the star can be
found with the help of the so-called Sobolev functions
\begin{equation}\label{eq:Toy3}
u_\alpha(r)\equiv\frac{(\alpha R)^{1/2}}{\big[r^2+(\alpha R)^2\big]^{1/2}},
\end{equation}
which satisfy
\begin{equation}\label{eq:Toy4}
\nabla^2u_\alpha=-3u_\alpha^5.
\end{equation}
Considering the function $Cu_\alpha$, we find that this function
satisfies Eq.~(\ref{eq:Toy1a}) for any choice of $\alpha$, given that
$C=(2\pi\rho_0/3)^{-1/4}$.  Indeed {\em any} solution $\bar\CF$ to
Eqs.~(\ref{eq:Toy1}) must be of this form in the interior of the star:
The function $Cu_{\bar\alpha}$ with $\bar\alpha=C^2[\bar\CF(0)]^{-2}$
has the same value and derivative as $\bar\CF$ at the origin, and as
we show in the next section, this implies that $\bar\CF\equiv
Cu_{\bar\alpha}$ throughout the interior of the star.

In the exterior, the only solutions of the flat-space Laplace
equation with asymptotic value unity are the functions $\beta/r+1$,
for some parameter $\beta$.  Consequently, any solution to
Eq.~(\ref{eq:Toy1}) must be a member of the family of functions
\begin{equation}\label{eq:Toy4_qq}
\CF(r)=\left\{\begin{aligned}&Cu_{\alpha}(r),&r<R\\
&\frac{\beta}{r}+1,&r>R
	      \end{aligned}
\right.
\end{equation}
with $C$ given above, and $\alpha$, $\beta$ real parameters.  The
parameters $\alpha$ and $\beta$ are determined by continuity of $\CF$
and its first derivative at the surface of the star,
\begin{align}
\frac{\beta}{R}+1 &= C u_\alpha(R),\label{eq:Toy4a}\\
-\frac{\beta}{R^2} &= C u'_\alpha(R),\label{eq:Toy4b}
\end{align}
where a prime denotes $\partial/\partial r$.  Eliminating $\beta$, we
find that $\alpha$ has to satisfy,
\begin{equation}\label{eq:Toy5}
\rho_0 R^2=\frac{3}{2\pi}f^2(\alpha),
\end{equation}
where 
\begin{equation}
f(\alpha)=\frac{\alpha^5}{(1+\alpha^2)^3}.
\end{equation}
Given a value for $\alpha$ we can find $\beta$ from (\ref{eq:Toy4a})
or (\ref{eq:Toy4b}), which then completely specifies a solution to
(\ref{eq:Toy1}).

The non-uniqueness of the solutions arises through the properties of
the function $f(\alpha)$.  We can see immediately that $f(\alpha)$
approaches zero for both $\alpha \to 0$ and $\alpha \to \infty$.  For
a sufficiently small value of $\rho_0 R^2$ in (\ref{eq:Toy5}) we may
therefore pick either a small or a large value of $\alpha$, which, as
we will show below, corresponds to either a strong-field or a
weak-field solution.

Examining $f(\alpha)$ more carefully, we see that it takes its maximum
at $\alpha_c=\sqrt{5}$.  Therefore, Eq.~(\ref{eq:Toy5}) has no
solution if $\rho_0$ is larger than the critical value
\begin{equation} \label{eq:rho_crit}
\rho_c=\frac{3}{2\pi R^2}f^2(\alpha_c)=\frac{3}{2\pi R^2}\frac{5^5}{6^6}
\approx \frac{0.0320}{R^2}.
%
\end{equation}
At the critical density Eq.~(\ref{eq:Toy5}) has exactly one solution,
$\alpha=\alpha_c$, while below the critical density there are two
solutions; one with $\alpha<\alpha_c$ and one with $\alpha>\alpha_c$.
This behavior is in complete analogy to the behavior of the extended
conformal thin-sandwich system examined in PY.

Having just derived all solutions to Eq.~(\ref{eq:Toy1}), we now now
discuss their properties in more detail.  It turns out to be convenient
to parametrize these solutions by $\alpha$.  Each value of $\alpha$
corresponds to precisely one solution with $\rho_0$ given by
Eq.~(\ref{eq:Toy5}).  Both limiting cases, $\alpha\to 0$ and
$\alpha\to\infty$ correspond to the limit of vanishing mass-density
(see Eq.~(\ref{eq:Toy5})).

We begin by computing the ADM-energy, which can be found using
Eq.~(\ref{eq:Toy4a}),
\begin{equation}
E=2\beta=\frac{2}{\alpha^2}R.
\end{equation}
For large $\alpha$, the ADM-energy tends to zero and we recover flat
space.  In the limit $\alpha\to 0$, however, the energy grows without
bound, despite the fact that $\rho\to 0$ as $\alpha\to 0$.  This
establishes the $\alpha > \alpha_c$ branch as the weak-field branch,
and $\alpha < \alpha_c$ as the strong-field branch.  We show a graph
of the energy as a function of density in Figure~\ref{fig:ToyEnergy}.

\begin{figure}
\includegraphics[scale=0.5]{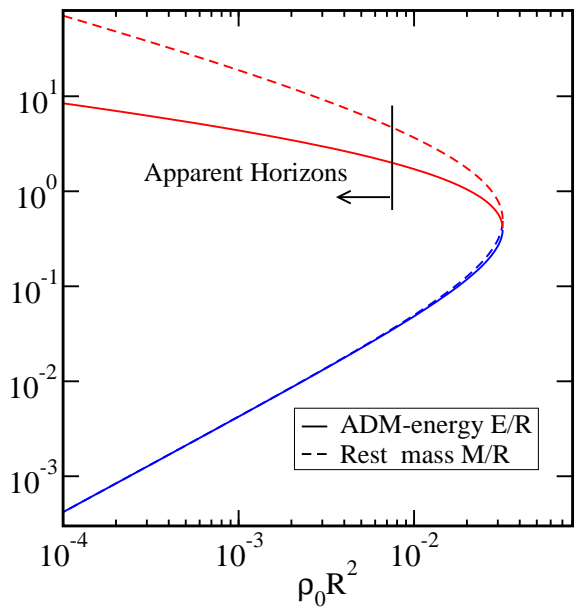}
\caption{\label{fig:ToyEnergy}ADM-energy and rest mass as function of
$\rho_0 R^2$ for the constant density star of Section
\ref{SubSec:ToyModel}.}
\end{figure}

Next we consider the rest mass $M$ of the star, which is given by
\begin{align}\nonumber
M&=\int_{r<R} \rho_0 \sqrt{g} dV=\int_0^R \rho_0 \CF^6\, 4\pi r^2\,dr\\
&=\frac{3}{4\alpha^5}\left[\alpha-\alpha^5+(1+\alpha^2)^3\arctan(\alpha^{-1}
)\right]R,
\end{align}
where we have used Eq.~(\ref{eq:Toy5}) to eliminate $\rho_0$.  This
expression has the limiting values
\begin{align}
M&\approx \frac{3\pi}{\alpha^5}R &&\mbox{for }\alpha\to 0,\\
M&\approx \frac{2}{\alpha^2}R &&\mbox{for }\alpha\to\infty.\label{eq:M+}
\end{align}
The weak-field limit $\alpha\to\infty$ corresponds to the limit in
which the star has vanishing mass, whereas the strong-field limit
$\alpha\to 0$ results in a star with unbounded mass, even though the
density itself approaches zero.  This behavior is caused by the fact
that the conformal factor, and hence the proper volume inside the
stellar radius $R$ diverges more rapidly than the rate at which the
density $\rho_0$ vanishes.

We point out that for all $\alpha>0$ we have $E<M$, so that the star
has negative binding energy as expected.  In the Newtonian limit
$\alpha\to\infty$ we recover the Newtonian binding energy,
\begin{equation}
E-M\approx -\frac{12}{5\alpha^4}R\approx-\frac{3}{5}\frac{M^2}{R},
\end{equation}
where we have used Eq.~(\ref{eq:M+}) in the second step.

Finally, we locate the apparent horizons in this family of initial
data sets.  For a time-symmetric hypersurface, apparent horizons
coincide with maximal surfaces, which in spherical symmetry and
conformal flatness are given by the roots of
\begin{equation}
\frac{\partial\CF}{\partial r}+\frac{\CF}{2r}=0.
\end{equation}
For $\alpha\!>\!1$, no roots to this equation exist, so that the
initial data surface does not contain an apparent horizon.  For
$\alpha<1$, two roots exist, one in the interior of the star at
$r=\alpha R$, and one in the exterior at $r=R/\alpha^2$.  The latter
one is the outermost extremal surface, which is the apparent horizon.
Both extremal surfaces merge on the surface of the star for
$\alpha=1$.  The density, ADM-energy and rest mass at formation of the
apparent horizon are $\rho_0|_{\alpha=1}=3/(128\pi R^2)\approx 0.0075
R^{-2}$, $E=2R$, and $M|_{\alpha=1}=3\pi R/2\approx 4.71 R$.

We now turn our attention to the critical point.  Around its maximum
$f(\alpha)$ behaves like a parabola, therefore 
\begin{equation}\label{eq:Parabola}
\alpha-\alpha_c\propto
 \pm(\rho_c-\rho_0)^{1/2}.
\end{equation}
Energy and mass at the critical point are
\begin{align}
E_c&=\frac{2}{5}R,\\
M_c&=\frac{18}{125}\left(9\sqrt{5}\arctan(1/\sqrt{5})-5\right)R\approx0.499R
,
\end{align}
respectively.  Since $\partial E/\partial\alpha\neq 0$ there, the
energy also changes parabolically with $\rho_0$,
\begin{equation}
E-E_c\;\propto\;\alpha-\alpha_c\;\propto\;
 \pm(\rho_c-\rho_0)^{1/2}.
\end{equation}
This parabolic behavior is apparent in Fig.~\ref{fig:ToyEnergy}.

At the critical point the local uniqueness of solutions must break
down, since the two branches meet there.  For this to happen, the
linearized operator must have a non-trivial solution at the critical
point.  The linearization of Eq.~(\ref{eq:Toy1}) reads
\begin{subequations}
\label{eq:Toy7}
\begin{equation}\label{eq:Toy7a}
\nabla^2\delta\CF+10\pi\rho_0\CF^4\,\delta\CF=0,
\end{equation}
with boundary conditions
\begin{align}\label{eq:Toy7b}
\frac{\partial\delta\CF}{\partial r}=0,& \qquad r=0\\
\delta\CF\to 0,& \qquad r\to\infty.\label{eq:Toy7c}
\end{align}
\end{subequations}

We will now construct all solutions of Eq.~(\ref{eq:Toy7}).  While
doing so, we consider $\rho_0$ as given and fixed.  If $\delta\psi=0$
is the only solution, then the kernel of this equation is trivial, and
solutions to the non-linear equation (\ref{eq:Toy1}) are locally
unique.  As just argued, at the critical point this will not be the
case, and there must be a non-zero solution of Eq.~(\ref{eq:Toy7}).
As it turns out, we can construct this solution analytically.

The key to solving Eq.~(\ref{eq:Toy7}) are again the Sobolev functions
$u_\alpha$.  Recall that $\CF=Cu_\alpha$ satisfies Eq.~(\ref{eq:Toy1})
in the interior of the star for {\em any} value of $\alpha$.  We can
therefore take the derivative of Eq.~(\ref{eq:Toy1}) with respect to
$\alpha$ and find
\begin{equation}\label{eq:Toy8}
\nabla^2\frac{\partial
u_\alpha}{\partial\alpha}+10\pi\rho_0 C^4u_\alpha^4\frac{\partial
u_\alpha}{\partial \alpha}=0.
\end{equation}
Choosing $\alpha$ to be a solution of (\ref{eq:Toy5}), so that it is
consistent with $\rho_0$, we can identify $C^4u_\alpha^4=\CF^4$, and
Eq.~(\ref{eq:Toy8}) reduces to Eq.~(\ref{eq:Toy7a}).  Consequently,
any function $A \partial u_\alpha/\partial\alpha$, with $\alpha$ given
by (\ref{eq:Toy5}) and $A$ an arbitrary constant, satisfies
Eq.~(\ref{eq:Toy7a}).  This forms a one-parameter family of functions,
{\em all} of which automatically satisfy the differential equation
Eqs.~(\ref{eq:Toy7a}) and the boundary condition (\ref{eq:Toy7b}) in
the interior.

Solutions $\delta\CF$ in the exterior must satisfy the Laplace
equation and the outer boundary condition (\ref{eq:Toy7c}), i.e. they
must take the form $B/r$ for some constant $B$.  Since this is a
one-parameter family of solutions, we have found {\em all} solutions
to Eqs.~(\ref{eq:Toy7a}) and (\ref{eq:Toy7c}) in the exterior.

To find a global solution $\delta\psi$ we now have to find constants
$A$ and $B$ so that the interior solution $A\partial
u_\alpha/\partial\alpha$ matches the exterior solution $B/r$
continuously in both the functions and their first derivatives at the
stellar radius $r = R$.  As expected, non-trivial solutions with
non-zero $A$ and $B$ exist only at the critical point
$\alpha=\alpha_c$.  There, the solution $\delta \psi$ takes the form
\begin{equation}
\delta\CF_c(r)\propto\left\{\begin{aligned}
&\frac{5R^2-r^2}{(r^2+5R^2)^{3/2}},&r<R\\ &\frac{4}{6^{3/2}r},&r>R
		    \end{aligned}
\right.
\end{equation}
and it is easy to verify that it indeed satisfies Eq.~(\ref{eq:Toy7})
at the critical point.

\subsection{Results for general \boldmath $\rho\ge 0$}
\label{SubSec:Math}

A fully worked out example like the constant density star presented
above is very instructive.  However, the example itself does not
provide any indication whether its behavior is generic.  In this
Section we prove theorems valid for general $\rho\ge 0$ with compact
support, indicating that the behavior found for the constant density
star is indeed generic for the Hamiltonian constraint with unscaled
matter density.  We will first show that for sufficiently ``large''
matter-densities $\rho$, no solution exists. We will then consider the
critical point and show that if a critical point exists, the solution
must vary parabolically close to it, as did the constant density star,
cf. Eq.~(\ref{eq:Parabola}).  Finally, we will prove a result which was
stated above to show that all solutions to the constant density star
have been found: If two functions each satisfying
Eqs.~(\ref{eq:Toy1a}) and (\ref{eq:Toy1b}) have the same value at the
origin, then they are identical.



We start with some preliminaries.  Rewriting the Laplacian in
Eq.~(\ref{eq:Toy1}) we find
\begin{equation}
\left(\CF+r\CF'\right)'=-2\pi r\rho\CF^5\le 0.
\end{equation}
Therefore the combination $\CF+r\CF'$ is monotonically decreasing and
bounded from below by its asymptotic value for large $r$,
\begin{equation}\label{eq:Math1}
\CF+r\CF'\ge 1.
\end{equation}
Furthermore, integrating Eq.~(\ref{eq:Toy1}) over a sphere of radius
$R$ we find 
\begin{equation}\label{eq:Math2}
4\pi R^2\; \CF'(R) = 
-2\pi \int_{R} \rho\CF^5 dV.
\end{equation}
Since $\rho\ge 0$ and $\CF>0$ we have $\CF'\le 0$, so that $\CF$ is a
decreasing function of radius, which is bounded from below by its
asymptotic value, $\CF\ge 1$.

We can now show that for sufficiently large $\rho$ Eq.~(\ref{eq:Toy1})
does not admit strictly positive solutions $\CF$.  Solving
Eq.~(\ref{eq:Math2}) for $\CF'(R)$ and substituting into
Eq.~(\ref{eq:Math1}) we find
\begin{equation}
2R\big[\CF(R)-1\big]\ge \int_{R} \rho\, \CF^5\, dV\ge\CF(R)^{5-k}
\int_R\CF^{k}\rho\,dV,
\end{equation}
for any $k\le 5$, where the last inequality follows from $\CF'\le 0$.
Rearranging terms we obtain
\begin{equation}
\frac{1}{R}\int_R\CF^k\rho\, dV\le 2\frac{\CF(R)-1}{\CF(R)^{5-k}}.
\end{equation}
The right hand side of this inequality is bounded independently of the
value of $\CF(R)$ by the biggest value of the function
\begin{equation}
g_k(x)=2\frac{x-1}{x^{5-k}}
\end{equation}
for $x\ge 1$.  For $k\le4$, this function is bounded by
\begin{equation}
g_k(x)\le C_k \equiv 2\frac{(4-k)^{4-k}}{(5-k)^{5-k}},
\end{equation}
so that any solution $\CF$ satisfies the integral bounds 
\begin{equation}\label{eq:MathBound}
\frac{1}{R}\int_R \CF^k \rho\, dV\le C_k 
\end{equation}
for any $k \le 4$.  For a given $k$, the bound $C_k$ is independent of $R$.
For positive $k$, $0<k\le 4$, this inequality constrains how large
solutions can be.  For example,
\begin{equation}
C_4\ge \frac{1}{R}\int_R\CF^4\rho\,dV\ge \frac{\CF(R)^4}{R}\int_R\rho\, dV
\end{equation}
implies 
\begin{equation}
\CF(R)\le C_4^{1/4}\left(\frac{1}{R}\int_R\rho\,dV\right)^{-1/4},
\end{equation}
which is a bound of how quickly the ``upper'' branch can diverge as
$\rho\to 0$.

 For $k=0$, the inequality~(\ref{eq:MathBound}) becomes independent of $\CF$: If a solution
$\CF$ exists for a certain $\rho$, then
\begin{equation}\label{eq:MathBound1}
\frac{1}{R}\int_R\rho\, dV\le 2 \frac{4^4}{5^5}\approx 0.163. 
\end{equation}
Equation~(\ref{eq:MathBound1}) holds for any $R$ for any strictly
positive solution of Eq.~(\ref{eq:Toy1}), therefore if a density
distribution $\rho(r)$ satisfies
\begin{equation}\label{eq:MathBound2}
m(R) \equiv \int_{R} \rho\,dV > 2\frac{4^4}{5^5}R
\end{equation}
even for one $R$, then no regular solution to the Hamiltonian
constraint Eq.~(\ref{eq:Toy1}) exists for this density.

For the constant density star, $m(r)$ is largest at the surface of the
star, $r=R$, where
\begin{equation}
m(R)=\rho_0\int_R 4\pi r^2 dr=\frac{4\pi}{3}\rho_0R^3.
\end{equation}
Equation~(\ref{eq:MathBound2}) then gives the necessary bound
$\rho_0\lesssim 0.0389/R^2$ for the existence of solutions.
Comparison with the exact critical density $\rho_c=0.0320/R^2$ from
Eq.~(\ref{eq:rho_crit}) reveals that the upper bound of the theorem is
only 20 per cent larger than the exact critical density (see also
Fig.~\ref{fig:ToyEnergy}).

Let us now examine the character of the critical point.  We take a
smooth sequence of non-negative densities, $\rho_{\gamma}$, such that
$\rho \equiv 0$ when $\gamma = 0$.  We then look for a smooth sequence
of solutions $\CF_\gamma$ to Eq.~(\ref{eq:Toy1}) with the density
$\rho$ given by $\rho_\gamma$, starting from $\CF \equiv 1$ at
$\gamma=0$.  The Implicit Function Theorem tells us that as long as
the linearized equation, Eq.~(\ref{eq:Toy7}),
\begin{subequations}
\label{eq:lin}
\begin{equation}\label{eq:lina}
\nabla^2\delta\CF+10\pi\rho_{\gamma}\CF_{\gamma}^4\,\delta\CF=0,
\end{equation}
with boundary conditions
\begin{align}\label{eq:linb}
\frac{\partial\delta\CF}{\partial r}=0,& \qquad r=0\\
\delta\CF\to 0,& \qquad r\to\infty.\label{eq:linc}
\end{align}
\end{subequations}
has {\em no} non-trivial solution, then the full nonlinear equation
\begin{subequations}
\label{eq:beta}
\begin{equation}\label{eq:betaa}
\nabla^2\CF_{\gamma}+2\pi\rho_{\gamma}\CF_{\gamma}^5=0,\\
\end{equation}
with boundary conditions
\begin{align}
\frac{\partial\CF}{\partial r}=0,&\qquad r=0,\\
\CF_{\gamma}\to 1,&\qquad r\to\infty,
\end{align}
\end{subequations}
has a regular solution which changes smoothly as a function of $\gamma$.

The obvious question to ask is what happens if the sequence approaches
the point where the first kernel of Eq.~(\ref{eq:lin})
appears\footnote{Clearly there are sequences
$\rho_\gamma$ for which this never happens,
e.g. $\rho_\gamma\equiv0$.}. Let us assume this happens at $\gamma_0$.
The trick is to consider the limiting process rather than the limit
point itself.  We know that when $\gamma = \gamma_0$ the equation
\begin{equation}\label{eq:gs}
\nabla^2\theta + 10\pi\rho_{\gamma_0}\CF_{\gamma_0}^4\theta = 0,
\end{equation}
has a positive solution $\theta$, going to zero at infinity.  This is the ground state of a Schr\"odinger
equation, because it is the {\it first} appearance of a kernel, hence it has no nodes, and thus it can be chosen to be everywhere positive.  

  We now differentiate
Eq.~(\ref{eq:betaa}) with respect to $\gamma$ (at any $\gamma < \gamma_0$ to find
\begin{equation}\label{eq:beta'}
\nabla^2\frac{d\CF_{\gamma}}{d\gamma} +
10\pi\rho_{\gamma}\CF_{\gamma}^4\frac{d\CF_{\gamma}}{d\gamma} = 
- 2\pi\frac{d\rho_{\gamma}}{d\gamma}\CF_{\gamma}^5.
\end{equation}
Multiplying Eq.~(\ref{eq:gs}) by $d\CF_{\gamma}/{d\gamma}$, 
Eq.~(\ref{eq:beta'}) by $\theta$ and subtracting the results
we obtain
\begin{align}\label{eq:diff}
\nabla \cdot (\frac{d\CF_{\gamma}}{d\gamma}&\nabla \theta -
\theta \nabla \frac{d\CF_{\gamma}}{d\gamma}) = \frac{d\CF_{\gamma}}{d\gamma}\nabla^2 \theta -
\theta \nabla^2 \frac{d\CF_{\gamma}}{d\gamma}\nonumber \\
 =& 10\pi(\rho_{\gamma}\CF_{\gamma}^4 - \rho_{\gamma_0}\CF_{\gamma_0}^4)\frac{d\CF_{\gamma}}{d\gamma}\theta
 +
2\pi\frac{d\rho_{\gamma}}{d\gamma}\CF_{\gamma}^5\theta.
\end{align}

Next we wish to integrate this equation over the whole space.  Let us
assume that $\rho_{\gamma}$ has compact support, or, at least, falls
off rapidly at infinity.  Both $\theta$ and $d\CF_{\gamma}/d\gamma$
fall off at infinity like $1/r$ and their first derivatives fall off
like $1/r^2$. This means that the total divergence upon integration
becomes a surface term, and the integrand falls off like
$1/r^3$. Therefore the integral vanishes. We get
\begin{equation}\label{eq:int}
5\int(\rho_{\gamma}\CF_{\gamma}^4 - \rho_{\gamma_0}\CF_{\gamma_0}^4)\frac{d\CF_{\gamma}}{d\gamma}\,\theta\, dV +
\int \frac{d\rho_{\gamma}}{d\gamma}\CF_{\gamma}^5\,\theta\, dV = 0.
\end{equation}

Consider this in the limit as $\gamma \rightarrow \gamma_0$.  The
second term tends to the constant
\begin{equation}
{\cal I}\equiv
\int\left.\frac{d\rho_\gamma}{d\gamma}\right|_{\gamma_0}\!\CF^5_{\gamma_0}\,\theta\,
dV.
\end{equation}
Note that both $\theta$ and $\CF_{\gamma_0}$ depend only on
$\rho_{\gamma_0}$, and not its derivative.  Therefore, changing
$\left.d\rho_\gamma/d\gamma\right|_{\gamma_0}$ via
\begin{equation}
\rho_\gamma \to \rho_\gamma + v(\gamma-\gamma_0) 
\end{equation}
for any function $v(r)$ will change ${\cal I}$ as 
\begin{equation}
{\cal I}\to {\cal I}+\int v\, \CF^5_{\gamma_0}\,\theta\,dV.
\end{equation}
Clearly, except for special instances, ${\cal I}$ will be non-zero.
Let us now assume this generic case, ${\cal I}\neq 0$ .  In the limit
$\gamma\to\gamma_0$, the second term in Eq.~(\ref{eq:int}) becomes the
non-zero constant ${\cal I}$, whereas the first term seems to go to
zero.  This cannot be and thus we are forced to conclude that the
limiting process as $\gamma \rightarrow \gamma_0$ must be somehow
singular.

The only thing that can possibly go bad is that $d\CF_{\gamma}/d\gamma
\rightarrow \infty$. Not only has it to blow up, it must do so over an
extended region. This is the only way that the integral, in the limit,
can go to a nonzero value.  Let us assume
\begin{equation}
\frac{d\CF_\gamma}{d\gamma}\propto (\gamma_0-\gamma)^p
\end{equation}
for some negative power $p<0$.  This implies
\begin{equation}
\CF_\gamma-\CF_{\gamma_0} \propto (\gamma_0-\gamma)^{p+1}. 
\end{equation}

Consider now the term 
\begin{align}
\rho_\gamma\CF_\gamma^4&-\rho_{\gamma_0}\CF^4_{\gamma_0}
=\left(\rho_\gamma-\rho_{\gamma_0}\right)\CF_\gamma^4 \\ \nonumber
+&\rho_{\gamma_0}\left(\CF_\gamma^3+\CF_\gamma^2\CF_{\gamma_0}
+\CF^\gamma\CF^2_{\gamma_0}+\CF^3_{\gamma_0}\right)
\left(\CF_\gamma-\CF_{\gamma_0}\right)
\end{align}
in the first integral in Eq.~(\ref{eq:int}).  The first term on the
right hand side scales as $\gamma_0-\gamma$ as the critical point is
approached, whereas the second one scales as
$(\gamma-\gamma_0)^{p+1}$.  Because $p+1<1$, the second term
dominates, and the full integrand scales as
\begin{align}\label{eq:fullScaling}
\left(\rho_\gamma\CF_\gamma^4-\rho_{\gamma_0}\CF^4_{\gamma_0}\right)\frac{d\CF_\gamma}{d\gamma}
\propto (\gamma_0-\gamma)^{p+1}(\gamma_0-\gamma)^p
\end{align}
close to the critical point.  In the generic case, the integral has to
approach the finite, non-zero value ${\cal I}$ in the limit
$\gamma\to\gamma_0$, which can only happen if $p = -1/2$.  Therefore,
the solution must vary parabolically,
\begin{equation}
\CF_\gamma-\CF_{\gamma_0}\propto \left(\gamma-\gamma_0\right)^{1/2}.
\end{equation}
This is exactly the behavior we have seen
in the constant density star model and also with what Pfeiffer and
York \cite{Pfeiffer-York:2005} observed.

This parabolic nature of solutions near the critical point can be
demonstrated explicitly even in the non spherically symmetric case by
using Lyapunov-Schmidt techniques \cite{OMW}.  We conjecture that a
second branch of solutions exists beyond the critical point as a
consequence of this parabolic behavior.  We have seen this explicitly
for the constant density stars, and we again refer to \cite{OMW} for a
more general treatment.

So far we have shown that there are distributions $\rho(r)$ for which
no solutions of Eq.~(\ref{eq:Toy1}) exist, and based on the parabolic
nature of the solutions at the critical point we have conjectured that
there are distributions for which exactly two solutions exist.  We do
not know whether this is generic.

Having moved past the first critical point, an open question is
whether another critical point is reached. Immediately past the first
critical point, $\rho$ does not change, but the conformal factor
increases.  In the language of the Schr\"odinger equation this means 
that the potential deepens
and the zero-energy ground state becomes a bound state with negative energy.  
As one moves
away from the critical point along the upper branch one is moving
`back' toward smaller $\gamma$, and so we expect $\rho_{\gamma}$ to
decrease while $\CF$ continues to increase. One could have that $\rho
\CF^5$ increases enough that the first excited state appears with zero
energy, or that $\rho \CF^5$ decreases again so that the ground state
becomes a zero energy state again.  In either case, the system reaches
another critical point and the solution curve may turn again.
Alternatively, $\rho_\gamma\CF_\gamma^5$ may be such that neither of
these two cases happens and the solution continues on all the way to
$\gamma = 0$.  This last alternative occurs for the constant density
star, as we have shown by explicit calculation; however, we do not
know whether this behavior is generic.

Finally, we show that if we have two positive solutions $\CF_1$ and
$\CF_2$ to Eq.~(\ref{eq:Toy1}) whose maxima agree, then they are
identical.  To prove this we first note that the maxima of both must
occur at $r = 0$.  The maximum principle tells us that there cannot be
a positive minimum, therefore there can only be one maximum, and
therefore it must occur at the origin.  Therefore at $r=0$, the two
functions $\CF_1$ and $\CF_2$ agree, their first derivatives both
vanish, and the second derivatives are equal (from
Eq.(\ref{eq:Toy1})). By differentiating Eq.(\ref{eq:Toy1}), one can
show that all the derivatives of the two functions agree at $r =
0$. If the functions were analytic, we were done.  However, there is
no reason to expect that this be true.  We need a more subtle
argument.

Track $\CF_1$ and $\CF_2$ as they move out from the origin. If they remain
the same all the way to infinity, we are done.  Instead, let us assume that
at some point $\CF_1 > \CF_2$.  Therefore we must encounter a region in
which $d(\CF_1 - \CF_2)/dr > 0$ and $\CF_1 - \CF_2 > 0$ and inside this
region $\CF_1 \ge \CF_2$. Take a point in this region, call it $R_0$.
Consider the equations satisfied by $\CF_1$ and $\CF_2$, subtract one from
the other and integrate over the ball of radius $R_0$. We get
\begin{equation}
\int_{R_0}\nabla^2(\CF_1 - \CF_2) dV + 2\pi\int_{R_0} \rho(\CF_1^5 -
\CF_2^5) dV = 0.
\end{equation}
The Laplacian becomes a boundary term, which is positive, because the
gradient of the difference is positive at $R_0$, while the bulk term
is also non-negative. This cannot be, so therefore the initial
assumption that the functions are different must be incorrect.

\section{Summary and Discussion}
\label{Sec:Sum}

In this paper we investigate the reason for non-uniqueness in the
extended conformal thin sandwich equations~\cite{Pfeiffer-York:2005}.
We argue that a term with the ``wrong sign'' in the elliptic equation
determining the lapse, Eq.~(\ref{eq:Lapse3}), is the cause for
non-uniqueness.  The sign of this particular term is such that the
maximum principle cannot be applied to prove local uniqueness of
solutions.  We support our claim by examining a simpler equation
having a term with an analogous ``wrong sign'', namely the the
Hamiltonian constraint with unscaled~\cite{York:1979} matter source
$\rho$ (cf. Eq.~(\ref{eq:Ham2})).  Specializing to constant density
stars we construct analytical solutions.  We find two branches of
solutions -- a weak-field and a strong-field branch -- with properties
that are remarkably similar to those found by PY.

We comment briefly that solutions to the {\em original} conformal
thin-sandwich decomposition, consisting of the Hamiltonian constraint
(\ref{eq:Ham2}) and the momentum constraint (\ref{eq:Mom2}) only, are
unique \cite{Cantor:1977,Cantor-Brill:1981,York:1999,%
Bartnik-Isenberg:2004,Maxwell:2005,Pfeiffer-York:2005}.  PY found
multiple solutions only for the {\em extended} conformal thin-sandwich
decomposition, which includes the lapse equation (\ref{eq:Lapse2}) in
addition to the two constraints.  This, too, suggests that the
non-uniqueness is caused by the lapse equation, in accordance with our
findings.

Our findings are certainly relevant for numerical work: If one wants
to solve the extended conformal thin sandwich equations, then
apparently, the possibility of finding two solutions is unavoidable.
Whether this will pose a problem for numerical work is less clear.
Sufficiently far away from the critical point, the solutions along the
upper and lower branch are significantly different, and it should be
obvious which solution is desired (generally the ``lower'' one, which
reduces to flat space for trivial free data).  Many different
researchers have solved the extended conformal thin sandwich equations
without
problems~\cite{Baumgarte-Cook-etal:1997,Bonazzola-Gourgoulhon-Marck:1999,%
Uryu-Eriguchi:2000,Gourgoulhon-Grandclement-etal:2001,%
Gourgoulhon-Grandclement-Bonazzola:2001a,%
Grandclement-Gourgoulhon-Bonazzola:2001b,Yo-Cook-etal:2004,%
Cook-Pfeiffer:2004,Caudill-Cook-etal:2006,Baumgarte-Skoge-Shapiro:2004,%
Taniguchi-Baumgarte-etal:2005} and have obtained a solution with
satisfactory properties.  However, past success is no guarantee for
future success, and for choices of free data which may be interesting
in the future, non-uniqueness issues could very well arise, especially
if one is interested in solutions which happen to be ``close'' to the
critical point.  Indeed, in constrained evolutions schemes, which
solve elliptic equations similar to the extended conformal thin
sandwich equations, it was reported that the elliptic solver failed to
converge in near-critical collapse of Brill
waves~\cite{Choptuik-Hirschmann-Liebling-Pretorius:2003,Rinne:2005}.
It was further argued that this failure related to the ``wrong sign''
in a term of the maximum slicing
condition~\cite{Rinne:2005,Rinne-Steward:2005}.  How precisely a
numerical code behaves in such cases depends very sensitively on its
implementation.  Some algorithms may not converge at all, like the
multi-grid schemes
in~\cite{Choptuik-Hirschmann-Liebling-Pretorius:2003,Rinne:2005},
while other algorithms may converge to one of the two solutions
(e.g. the Newton-Raphson method used in PY;
cf. \cite{Pfeiffer-Kidder-etal:2003}).  In the former case it may be
difficult to ascertain whether failure of the numerical method is
indeed due to non-uniqueness properties of the underlying analytic
problem (rather than a bug), whereas in the latter case one is faced
with the question of which of the two solutions one wants, and how to
ensure convergence toward the desired solution.

\acknowledgments

We would like to thank Edward Malec and Darragh Walsh for helpful
comments, as well as the Isaac Newton Institute and the California
Institute of Technology for hospitality during various stages of this
work.  This research was supported in part by a grant from the Sherman
Fairchild Foundation, by NSF grant PHY-0601459 and by NASA grant
NNG05GG52G to Caltech, as well as by NSF Grant PHY-0456917 to Bowdoin
College.


\end{document}